\documentclass[12pt]{iopart}
\usepackage[utf8]{inputenc}

\expandafter\let\csname equation*\endcsname\relax
\expandafter\let\csname endequation*\endcsname\relax
\usepackage[ngerman,english]{babel}
\usepackage{amsmath} % Required for some math elements 
\usepackage{cmap}
\usepackage{iopams}
\usepackage{amsopn}
\usepackage{amssymb}
\usepackage{lmodern}
\usepackage{braket}
\usepackage{float}
\usepackage{graphicx} % Required for the inclusion of images
\usepackage{hyperref}
\usepackage{calligra}
\usepackage[numbers]{natbib}
\usepackage{doi}
\usepackage{xspace}
\usepackage{setspace}
\hypersetup{colorlinks=true, linkcolor=blue, citecolor=blue, urlcolor=blue}
\renewcommand{\i}{\mathrm i}
\renewcommand{\d}{\mathrm d}
\newcommand{\newblock}{}
\begin{document}

\title[Quantum coherent diffractive imaging]{Quantum coherent diffractive imaging}

\author{Björn Kruse$^{1}$, Benjamin Liewehr$^1$, Christian Peltz$^1$ and Thomas Fennel$^{1,2,*}$}
\address{$^1$Institute of Physics, University of Rostock, Albert-Einstein-Straße 23, D-18059 Rostock, Germany}
\address{$^2$Max-Born-Institut, Max-Born-Straße 2A, D-12489 Berlin, Germany}
\eads{\mailto{\color{blue}thomas.fennel@uni-rostock.de}}

%\vspace{10pt}
%\begin{indented}
%\item[]February 2014
%\end{indented}

\begin{abstract}
Coherent diffractive imaging has enabled the structural analysis of individual free nanoparticles in a single shot and offers the tracking of their light induced dynamics with unprecedented spatial and temporal resolution. The retrieval of structural information from scattering images in CDI experiments is so far mostly based on the assumption of classical scattering in linear response. For strong laser fields and in particular for resonant excitations, both the linear and the classical description may no longer be valid as population depletion and stimulated emission become important. Here we develop a density-matrix-based scattering model in order to include such quantum effects in the local medium response and explore the transition from linear to non-linear CDI for the resonant scattering from Helium nanodroplets. We find substantial departures from the linear response case already for experimentally reachable pulse parameters and show that non-linear spatio-temporal excitation dynamics results in rich features in the scattering, leading to the proposal of quantum coherent diffractive imaging (QCDI) as a promising novel branch in strong-field XUV and x-ray physics. 
\end{abstract}

%\maketitle
\maketitle

\section{Introduction}
Single-shot coherent diffractive imaging (CDI) has enabled the ultrafast characterization of the structure and dynamics of individual isolated nanoparticles in free flight~\cite{ChapmanNature2011,BostedtPRL2012,GorkhoverPRL2012,GorkhoverNatPhoton2018}. This unique capability has led to various fundamental scientific findings, including the shape analysis of viruses~\cite{SeibertNature2011}, tracking of laser-driven damage in clusters on femtosecond time scales~\cite{GorkhoverNatPhoton2016}, the discovery of previously unresolved metastable shapes in the growth process of metal clusters~\cite{BarkeNatComm2015}, and the direct visualization of quantum vortices in superfluid helium nanodroplets~\cite{GomezScience2014}, to name only a few.

The intense, ultrashort short-wavelength pulses required to record meaningful single-shot scattering patterns are typically only available at XUV and X-ray free electrons lasers (FELs)~\cite{EmmaNatPhoton2010,AllariaNatPhoton2013}. While CDI with hard x-rays for maximal spatial resolution is likely to remain the domain of XFELs, lab-based sources have become sufficiently brilliant to enable CDI in the XUV domain. Despite lower nominal spatial resolution, XUV-CDI enables the measurement of wide-angle scattering and thus provides three-dimensional structural information~\cite{BarkeNatComm2015}, which is particularly valuable for the characterization of non-reproducible targets~\cite{LangbehnPRL2018}. 

Only recently, lab-based single-shot CDI of single Helium nanodroplets could be demonstrated with a laser-driven table-top high-harmonic generation (HHG) source~\cite{RuppNatComm2017}. Besides marking a breakthrough regarding accessibility, laser-based sources open unprecedented conceptual opportunities for CDI applications. The most obvious one results from the exquisite temporal control of optical laser systems and promises extreme time-resolution that is likely to reach the sub-fs range for CDI with attosecond XUV pulses~\cite{SanderJPB2015}. The second aspect, which is relevant also for FELs if timing is less critical, is associated with the prospect of flexible multi-color CDI scenarios. Note that, despite limitations regarding pulse duration, seeded FELs can provide widely adjustable multi-color beams, even with accurate phase control~\cite{PrinceNatPhoton2016}. HHG-based multi-color scattering patterns have already been shown to enable the single-shot characterization of optical properties of non-reproducible targets~\cite{RuppNatComm2017}. Most importantly, because of the opportunity to provide intense pulses with durations comparable to the coherence times of XUV excitations, XUV-CDI with HHG sources (or appropriately seeded FELs) is expected to open unprecedented routes to monitor and temporally probe quantum coherent dynamics in excited nanostructures. 

So far, the analysis of CDI experiments is mostly based on the assumptions of classical scattering and a linear medium response. In this case the scattering process is described by solving Maxwell's equations using a linear complex refractive index~\cite{BarkeNatComm2015} or tabulated atomic scattering factors~\cite{SanderJPB2015} or by employing appropriate approximations such as the first Born approximation \cite{barty2013molecular}. 
However, for strong laser fields and especially for excitations near atomic resonances, the linear description may no longer be valid as effects like population transfer and depletion may become non-negligible. Specifically, the importance of quantum coherent dynamics, such as Rabi-cycling, stimulated emission, and induced transparency in the context of CDI are essentially unknown. Non-linear effects and transient changes of the optical response due to ionization have been included, e.g., via an effective damage-induced reduction of atomic form factors~\cite{ho2020role} or using explicit classical models for the resulting microscopic plasma motion~\cite{PeltzPRL2014}. Recent studies on the prospects of incoherent scattering from fluorescence and Compton scattering~\cite{schneider2018quantum,classen2017incoherent} and superfluorescence in the saturation regime~\cite{benediktovitch2019quantum} indicate the potential of using quantum effects in CDI scenarios. The so far missing firm understanding of the impact of spatio-temporal coherent bound-state dynamics is of paramount importance to (i) justify applicability of established CDI reconstruction methods employing linear scattering or (ii) to identify scenarios that enable to monitor associated non-linear processes, including the prospect of tracking the spatio-temporal excitation transport in extended quantum systems. 

Here we present a systematic exploratory theoretical analysis to address the above two central questions. We develop a consistent semi-classical model for the electromagnetic field propagation in the presence of quantum coherent bound state dynamics and apply the approach to the benchmark scenario of XUV scattering from Helium nanodroplets. Specifically, we solve Maxwell's equations in the time domain using a quantum mechanical local polarization response treated with a few-level density matrix model. The parameters of the model are matched to the linear optical properties, which we describe by a Drude-Lorentz reference model, ensuring convergence to the correct linear limit for low intensity. Our systematic analysis of the scattering from a sub-$\mu$m Helium droplet for resonant excitation of the 1s$^2$ to 1s2p transition shows severe departures from the linear response behavior for laser parameters that are feasible with available laser technology. Our theory predicts substantial inelastic wide-angle scattering features that are reminiscent of the Mollow triplet~\cite{MollowPhysRev1969} in atomic systems and are shown to reflect inhomogenous Rabi-Cycling and excitation waves. 

The remainder of this paper is structured as follows. In section~\ref{sec:Methods} we introduce the two employed local polarization models and their self-consistent solution along with the electromagnetic field-propagation. Section~\ref{sec:Results_ands_Discussion} contains a linear-response benchmark calculation to validate the few-level description and presents the main results for non-linear excitation. Finally, we summarize the main conclusions and provide an outlook for future work in section~\ref{sec_Conclusions}.

%\newpage
\section{Methods}
\label{sec:Methods}
The basis of our semi-classical scattering model is the explicit propagation of the electromagnetic field according to Maxwell's curl equations in the absence of free charges and for non-magnetic materials 
\begin{align}
\dot{{\boldsymbol{\cal B}}}&=-\nabla \times \mathbf{{\boldsymbol{\cal E}}} \label{eq:Maxwell1}\\
\dot{{\boldsymbol{\cal E}}} &= \frac{1}{\varepsilon_0 \mu_0}\nabla \times \boldsymbol{\cal B}-\frac{1}{\varepsilon_0}\dot{\mathbf{P}}\label{eq:Maxwell2}.
\end{align}
Here $\boldsymbol{\cal E}$ and $\boldsymbol{\cal B}$ are the spatially averaged electric and magnetic fields (continuum picture), $\varepsilon_0$ and $\mu_0$ are the permittivity and permeability of free space, and $\mathbf{P}$ is the polarization. The latter has to be propagated self-consistently with the fields, contains the complete medium response, and is non-zero only inside the target.
The equations are solved in 3D for a rectangular arena employing the Finite-Differences-Time-Domain method (FDTD)~\cite{TafloveFDTDbook2005} and using Yee staggering of all discretized fields~\cite{YeeIEEE1966}. 

Two models are used to describe the time-dependent electronic polarization response, i.e., (i) a standard classical linear response model and (ii) a quantum model that is based on a few-level density matrix description of the electron dynamics. 

\subsection{Drude-Lorentz model}
As the standard linear response reference we employ the Drude-Lorentz model (DLM) that mimics the local dispersive medium response via a set of damped harmonic oscillators. For a single oscillator per atom and an isotropic response, the atomic electric dipole ${\bf p}(t)$ is described by the classical equation of motion 
\begin{align}
\left( \frac{\partial^2}{\partial t^2} + \nu \frac{\partial}{\partial t} + \omega_0^2\right) {\bf p}(t) =f \frac{q_e^2}{{m_e}}{\boldsymbol{\cal E}}(t),
\label{eq:DLM1}
\end{align} 
where ${\boldsymbol{\cal E}}(t)$ is the electric field at the location of the oscillator, $\omega_{0}$ and $\nu$ are the oscillator's eigenfrequency and the associated collision frequency, $q_e$ and ${m_e}$ are the electron's charge and mass, and  $f$ is the oscillator strength. For an atomic number density $n_0$ the associated polarization ${\bf P}(\omega)=n_0 {\bf p}(\omega)$ in the Fourier domain ($\frac{\partial}{\partial t}\rightarrow -\i\omega $) becomes 
\begin{align}
{\bf P}(\omega)=\varepsilon_0 \chi(\omega) {\boldsymbol{\cal E}}(\omega)\qquad \mbox{with} \qquad \chi(\omega)=\frac{\omega_p^2}{\omega_{0}^2 - \omega^2-\i\omega \nu},
\label{eq:DLM2}
\end{align}
where $\omega_p=\sqrt{\frac{q_e^2 n_0}{\varepsilon_0 {m_e}}}$ is the plasma frequency. Although we do not consider  free charges, the plasma frequency can be used as a measure for the density to preserve the analogy to the Drude formula. The resulting relative permittivity $\varepsilon_r(\omega)=1+\chi(\omega)$ finally establishes the link to the refractive index 
\begin{align}
n(\omega)=\sqrt{\varepsilon_r(\omega)}=\sqrt{1+\chi(\omega)}=1-\delta(\omega)+\i\beta(\omega),
\label{eq:DLM3}
\end{align}
where $\delta$ and $\beta$ describe its real and imaginary parts.
A desired spectral dispersion can now be mimicked by a certain set of oscillator parameters ($\omega_0$, $\nu$, $f$). The generalization to multiple fractional oscillators (multiple resonances) is straightforward by representing the susceptibility as a sum $\chi(\omega)=\sum_i \chi_i(\omega)$, where each term reflects the response of an individual fractional oscillator according to Eq.~\eqref{eq:DLM2}. Note that this widely used superposition of uncoupled oscillators provides a valid description of the time-domain polarization only in the limit of linear response. For a generalization to a non-linear Lorentz model see Refs.~\cite{VarinCPC2018,VarinOME2019}. 

In this work we consider a single oscillator in the DLM and solve the equation of motion (Eq.~\eqref{eq:DLM1}) for the field-driven local dipole response in each cell self-consistently with the field evolution in Eqs.~\eqref{eq:Maxwell1}-\eqref{eq:Maxwell2}, where the local polarization velocity
\begin{align}
\dot {\bf P}_{\rm DLM}(t)=n_0\dot {\bf p}(t)
\label{eq:DLM4}
\end{align}
establishes the feedback to the field dynamics. This approach corresponds to the well-known auxiliary differential equation (ADE) method for the description of dispersion~\cite{TafloveFDTDbook2005}.

\subsection{Few-level density-matrix model}
\begin{figure}[h]
	\begin{center}
		\includegraphics[width=0.8\textwidth]{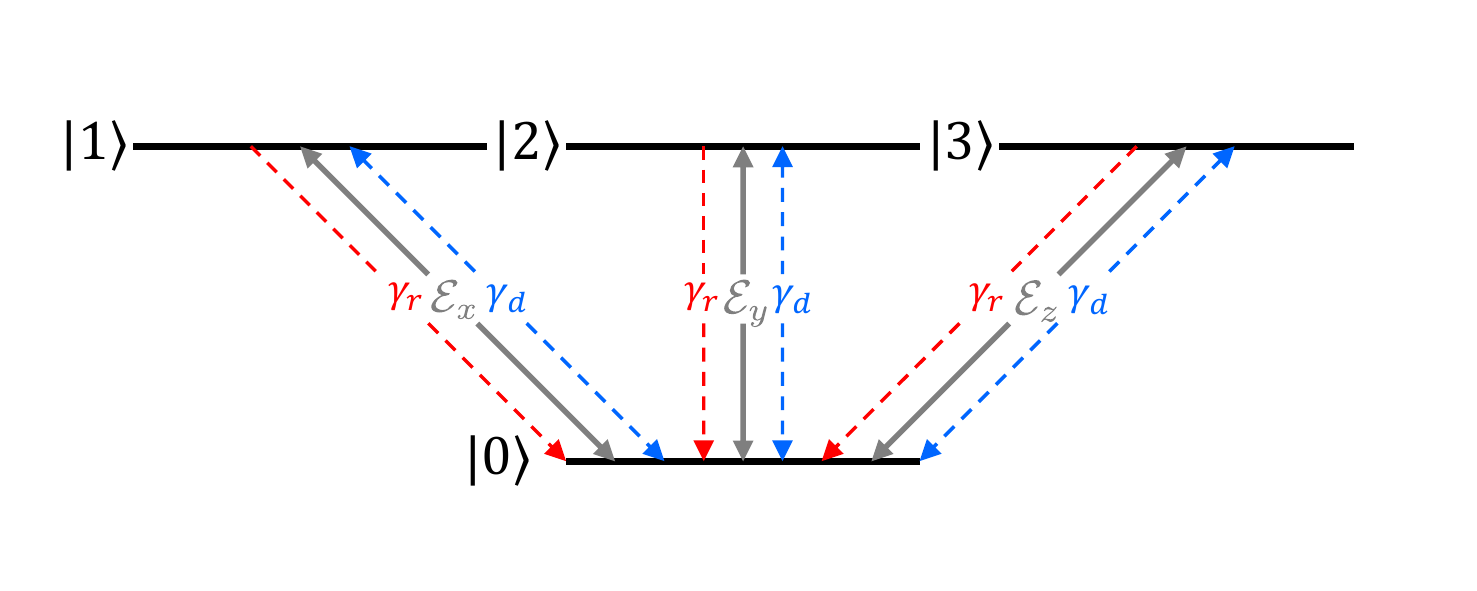}
		\caption{Schematic sketch of the active states of the employed few-level system to describe the Helium 1s$^2$-1s2p excitation. Here $|0\rangle$ is the ground state and $|1\rangle$, $|2\rangle$, and $|3\rangle$ are the three degenerate excited state orbitals (2p). Solid arrows indicate the coupling to the respective field components and dashed arrows indicate relaxation ($\gamma_r$) and decoherence ($\gamma_d$).}
		\label{fig:FLSskizze}
	\end{center}
\end{figure}

For the quantum mechanical description of the local response including coherence and non-linearity we use a few-level density-matrix (FLDM) model that is sketched in Fig.~\ref{fig:FLSskizze}.
The active Hilbert space is constructed from a ground state $|0\rangle$ with energy $E_0$ that reflects He 1s$^2$ and the threefold degenerate first excited 1s2p-states ($|1\rangle, |2\rangle, |3\rangle$) with energies $E_1=E_2=E_3$. Within the atomic orbital representation,  the dipole selection rules associated with the 1s-2p transition lead to a particularly simple structure of the matrix elements ${\boldsymbol \mu}_{ij}=\langle i | \hat {\boldsymbol \mu} | j \rangle$ of the dipole operator in the Schrödinger picture
\begin{align}
\hat {\boldsymbol \mu}=q_e\,\hat {\bf r}= \left[ \begin{array}{c c c c}
0               & \mu\,{\bf e}_x  & \mu\,{\bf  e}_y   & \mu\,{\bf  e}_z \\
\mu\,{\bf e}_x  & 0 & 0 & 0 \\
\mu\,{\bf e}_y  & 0 & 0 & 0 \\
\mu\,{\bf e}_z  & 0 & 0 & 0
\end{array}
\right ],
\end{align}
where $q_e$ is the charge of the electron, $\mu$ is the magnitude of the (isotropic) transition dipole moment and ${\bf e}_{i}$ with ($i=x,y,z$) are the Cartesian unit vectors. The associated expectation value of the dipole for a given density operator $\hat \rho$ follows as
\begin{align}
\langle \hat {\boldsymbol \mu}\rangle ={\rm Tr} (\hat {\boldsymbol \mu}\hat \rho)=\mu \left( \begin{array}{c}
\rho_{01} + \rho_{10} \\
\rho_{02} + \rho_{20} \\
\rho_{03} + \rho_{30} 
\end{array}\right). 
\end{align}
The system Hamiltonian contains the unperturbed atomic part (with $\hat H^{(0)}|i\rangle=E_i |i\rangle $) and the interaction term in dipole approximation according to
\begin{align}
\hat H=\hat H^{(0)} \underbrace {- \hat {\boldsymbol \mu}\cdot \boldsymbol{\cal E}(t)}_{\hat H^{({\rm int})}}
=\left[ \begin{array}{c c c c}
E_{0}               & 0   &  0  & 0 \\
0 & E_{1} & 0 & 0 \\
0 & 0 & E_{2} & 0 \\
0 & 0 & 0 & E_{3}
\end{array}
\right ]
-\mu\left[ \begin{array}{c c c c}
0               &  {\cal E}_x  &  {\cal E}_y   &  {\cal E}_z \\
 {\cal E}_x  & 0 & 0 & 0 \\
 {\cal E}_y  & 0 & 0 & 0 \\
 {\cal E}_z  & 0 & 0 & 0
\end{array}
\right ].
\end{align} 
The time evolution of the density matrix is evaluated in the interaction picture to remove the trivial unperturbed dynamics from the propagation. Matrix elements of a given operator $\hat A$ in the Schrödinger picture are transformed into the interaction picture representation (henceforward indicated with a tilde) via
\begin{align}
\tilde A_{mn}=A_{mn} e^{\i\omega_{mn}t}
\end{align}  with the transition frequency $\omega_{mn}=\frac{E_m-E_n}{\hbar}$. In the evolution described by the generalized von-Neumann equation 
\begin{align}
\dot {\tilde \rho}_{nm}=-\frac{\i}{\hbar}\sum_k \left[ \tilde H^{\rm int}_{nk} \tilde \rho_{km}-\tilde \rho_{nk}\tilde H^{\rm int}_{km} \right]-\gamma_{nm}\left( \tilde \rho_{nm}-\tilde \rho_{nm}^{eq.}\right)
\end{align}
the last phenomenological dissipation term has been added to describe decoherence and relaxation towards the equilibrium state $\hat {\tilde \rho}^{\rm eq.}$. The latter we consider to reflect the fully occupied ground state. In the absence of interaction, the dynamics is determined only by relaxation and decoherence. We assume symmetric relaxation and decoherence behavior between ground and excited states with 
\begin{align}
{\bar \gamma} =\left[ \begin{array}{c c c c}
\gamma_{\rm r}        & \gamma_d & \gamma_d & \gamma_d  \\
\gamma_d & \gamma_{\rm r} & 0 & 0 \\
\gamma_d & 0 & 	\gamma_{\rm r}& 0\\
\gamma_d & 0  & 0& \gamma_{\rm r}
\end{array}
\right ] \qquad \mbox{and}  \qquad 
\hat {\tilde \rho}^{\rm eq.}=\left[ \begin{array}{c c c c}
1   & 0 & 0 & 0\\
0   & 0 & 0 & 0\\
0   & 0 & 0 & 0\\
0   & 0 & 0 & 0
\end{array}
\right ], 
\end{align}
where the rates for decoherence and relaxation are denoted as $\gamma_d$ and $\gamma_r$, respectively. It should be emphasized that our assumptions of an isotropic response and equivalent excited states requires vanishing decoherence among excited states. Further we assume that relaxation, e.g. due to radiative decay, is negligible on the time scales of the considered fs scattering dynamics and set $\gamma_r=0$. We like to note that in principle the inclusion of additional relaxation channels is straightforward. For the considered 1s2p excited state, the only additional possible channel would be the 1s2p to 1s2s relaxation \cite{mudrich2020ultrafast}. As our scenario is focused on the coherent dynamics, its impact is considered negligible as long as the relaxation time is large compared to the dephasing time. 

The response properties of the FLDM model are determined by the excitation energy $E_1-E_0=\hbar \omega_{10}$, the magnitude of the transition dipole $\mu$, and the decoherence rate $\gamma_d$. In analogy to the DLM description, an individual local few-level system is propagated in each FDTD cell using the corresponding electric field to solve the evolution for the polarization. The expectation value of the dipole (semi-classical picture) in the interaction picture representation has the form
\begin{align}
\langle {\boldsymbol \mu}\rangle ={\rm Tr} (\hat {\tilde {\boldsymbol \mu}}\hat {\tilde \rho})=\mu \left( \begin{array}{c}
\tilde \rho_{01}e^{\i\omega_{10}t} + \tilde \rho_{10}e^{\i\omega_{01}t} \\
\tilde \rho_{02}e^{\i\omega_{20}t} + \tilde \rho_{20}e^{\i\omega_{02}t} \\
\tilde \rho_{03}e^{\i\omega_{30}t} + \tilde \rho_{30}e^{\i\omega_{03}t} 
\end{array}\right) 
\end{align}
with the associated dipole velocity  
\begin{align}
\langle \dot{\boldsymbol \mu}\rangle=\frac{\d}{\d t} {\rm Tr} (\hat {\tilde {\boldsymbol \mu}}\hat{\tilde \rho})
=\mu 
\left( \begin{array}{c}
\i\omega_{10}\left({\tilde \rho}_{01}e^{\i\omega_{10}t}-{\tilde \rho}_{10}e^{i\omega_{01}t}\right) +\dot {\tilde \rho}_{01}e^{\i\omega_{10}t}
 +\dot {\tilde \rho}_{10}e^{\i\omega_{01}t}
\\
\i\omega_{20}\left({\tilde \rho}_{02}e^{\i\omega_{20}t}-{\tilde \rho}_{20}e^{i\omega_{02}t}\right) +\dot {\tilde \rho}_{02}e^{\i\omega_{20}t}
+\dot {\tilde \rho}_{20}e^{\i\omega_{02}t}
\\
\i\omega_{30}\left({\tilde \rho}_{03}e^{\i\omega_{30}t}-{\tilde \rho}_{30}e^{\i\omega_{03}t}\right) +\dot {\tilde \rho}_{03}e^{\i\omega_{30}t}
+\dot {\tilde \rho}_{30}e^{\i\omega_{03}t}
\end{array}\right). 
\end{align}
The resulting polarization velocity 
\begin{align}
\dot {\bf P}_{\rm FLDM}(t)=n_0 \langle \dot{\boldsymbol \mu}\rangle
\end{align}
serves as the feedback for the field propagation, in close analogy to the DLM case in Eq.~\eqref{eq:DLM4}.

\subsection{Definition and matching the model parameters}
\begin{figure}[h]
	\begin{center}
	\includegraphics[width=0.6\textwidth]{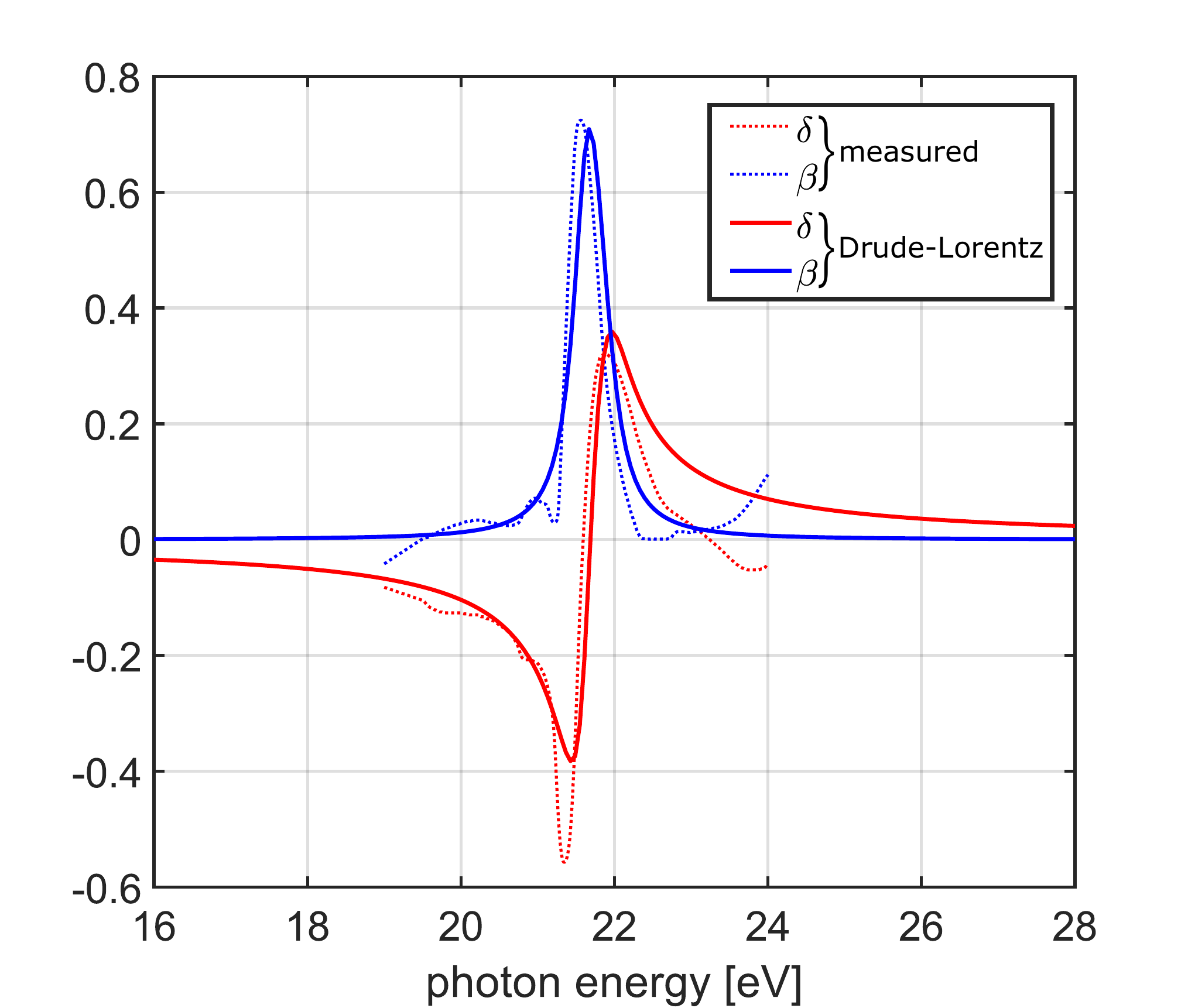}
	\end{center}
	\caption{Comparison of measured and fitted refractive index $n=1-\delta +\i \beta$ of bulk liquid helium in the linear regime in the vicinity of the helium 1s-2p transition~\cite{SurkoPRL1969,LucasPRB1983}. \label{fig:refrac_buil_He}}
\end{figure}

Both the DLM and the FLDM model describe the response properties via respective sets of oscillator parameters, where the DLM parameters are directly linked to the refractive index and the dispersion of the model system. In this work the response near the first excitation of Helium ($\mathrm{1s}^2 \rightarrow \mathrm{ 1s2p}$) will be modelled with a single oscillator, which requires to fix the three DLM parameters $\omega_0,\nu,f$. In order to describe liquid Helium we use parameters ($\hbar\omega_0=21.61 \:\mathrm{eV}$, $h \nu=0.43\:\mathrm{eV}$, and $f=0.49$) that where fitted to experimental data in Refs.~\cite{SurkoPRL1969,LucasPRB1983}. The resulting dispersion described by the DLM for an atomic density of $n_0=0.022 \:\text{\AA}^{-3}$ is compared to the experimental data from Refs.~\cite{SurkoPRL1969,LucasPRB1983} in Fig.~\ref{fig:refrac_buil_He}. At this point, all parameters of the DLM description are fixed. 

In order to determine the FLDM model parameters ($\gamma_d, \hbar \omega_{01}, \mu$), we require that the resulting linear optical properties must match the DLM description. Therefore the weak-field limit of the FLDM response has to be determined, which, because of isotropy of the linear response in our model, need to be considered only along one axis. e.g. the x-axis. The evolution of the expectation value of the dipole acceleration in the weak field limit ($\rho_{00}=1$ and $\rho_{jj}\approx 0$ for $j=1,2,3$) can then be written in a similar way as the classical oscillator equation Eq.~\eqref{eq:DLM1}
\begin{align}
\ddot {\langle \mu_x \rangle} +  \underbrace{2\gamma_d}_{\nu} \langle \dot \mu_x \rangle+ \underbrace{\left[\omega_{10}^2+\gamma_d^2\right]}_{\omega_0^2} \langle \mu_x \rangle  = \underbrace{\frac{2\omega_{10}\mu^2}{\hbar}}_{f\frac{q_e^2}{m_e}}  {\cal E}_x(t).
\end{align}
The expectation values of acceleration, velocity and magnitude of the dipole obviously coincide with the  classical result if the prefactors are matched as indicated. Solving the matching conditions for the FLDM parameters yields 
\begin{align}
\gamma_d=\frac{\nu}{2}, \qquad \omega_{10}=\sqrt{\omega_0^2-\frac{\nu^2}{4}}, \qquad \mbox{and} \qquad 
\mu=e\sqrt{\frac{\hbar f}{2\omega_{10}m_e}},
\end{align}
where $e$ is the elementary charge. The FLDM parameters ($\hbar \gamma_d =0.215\:\mathrm{eV}, \hbar\omega_{10}=21.61 \:\mathrm{eV}$, $\mu=0.2939~{\mathrm{e\AA}}=1.41\,\mathrm{D}$) are now fully matched to above DLM parameters.

\section{Results and Discussion}
\label{sec:Results_ands_Discussion}
\begin{figure}[h!]
	\begin{center}
		\includegraphics[width=0.75\textwidth]{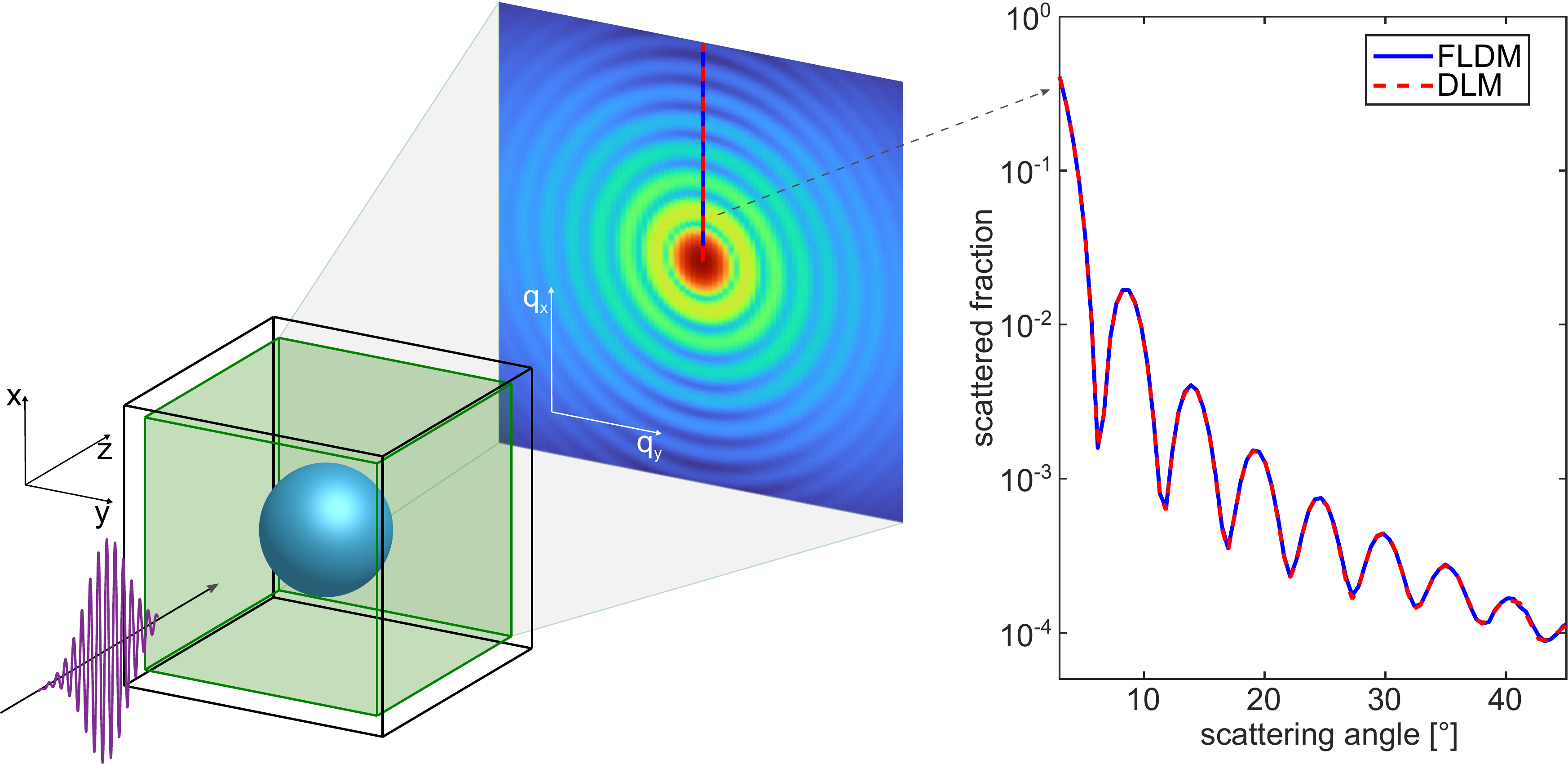}
		\caption{Sketch of the scattering scenario described in the simulations. A spherical helium droplet with radius $R=300\,{\rm nm}$ is placed in a cubic arena (left). The incident pulse with flat spatial intensity profile and Gaussian temporal profile propagates along the $z$-axis and is polarized along the $x$-axis. The equivalent source method, sampled on a control surface (green cube), is used to determine spectrally resolved far-field scattering images (as indicated).  Radial scattering intensity profiles (right) along the $q_x$ axis for $q_y=0$ for low incident field (linear response limit) are calculated using the FLDM method (blue line) and DLM (red line). \label{fig:scenario}}
	\end{center}
\end{figure}

To investigate the impact of the non-linear local polarization response on the scattering process and the final diffraction pattern for a realistic scenario, we considered a spherical Helium droplet with radius $R=300\,\mathrm{ nm}$ as a model system. The sphere is centered in the computational domain that is an equidistant, cubic grid of $251\times251\times251$ cells with a spatial resolution of $\Delta x = 2.8 \,\mathrm{nm}$, see Fig.~\ref{fig:scenario}. The resonant incident pulse with central photon energy of $\hbar\omega_\text{inc}=21.61\,\mathrm{eV}$ and flat phase front is linearly polarized along the $x$-axis, propagates along the $z$-axis and is described with a Gaussian temporal envelope with $8\:\mathrm{fs}$ duration (FWHM). Such a short duration is required to study coherent dynamics on the timescale of the dephasing time.

\subsection{Spectrally selective scattering images and linear response benchmark}
The incidence pulse is injected into the FDTD arena using the total-field-scattered-field scheme. The choice of the DLM or FLDM models determines only the engine that is used for the integration of the polarization response. The scattered fields are sampled on a cubic control surface (green) and decomposed into spectral components on-the-fly via a discrete Fourier transform. The spectral amplitudes and phases of the electric and magnetic fields on the control surface are used to construct electric and magnetic surface currents that represent an equivalent source~\cite{TafloveFDTDbook2005}. From the latter the scattering in the far-field is sampled for each spectral component, which yields spectrally-resolved scattering images. Note that this method allows the description of the scattering in full solid angle. Summation over individual spectral contributions in real space results in the final predicted scattering images, see Fig.~\ref{fig:scenario}. To verify consistence of our description for both polarization models, we calculated scattering images with the DLM and FLDM simulations for low pulse intensity. The comparison of the respective (spectrally integrated) scattering profiles in Fig.~\ref{fig:scenario} documents the agreement in the linear response limit and validates our method. The scattering images show the typical ring structure for spherical droplets~\cite{LangbehnPRL2018}.

\subsection{Population and electric field dynamics for non-linear excitaion}
Pronounced nonlinear features appear in the prediction of the FLDM model as the intensity of the incident pulse is increased. In the following we discuss as a representative case the excitation with pulse intensity $I=5\times10^{14}\:\mathrm{W/cm^2}$. 
\begin{figure}[h]
	\begin{center}
		\includegraphics[width=1.0\textwidth]{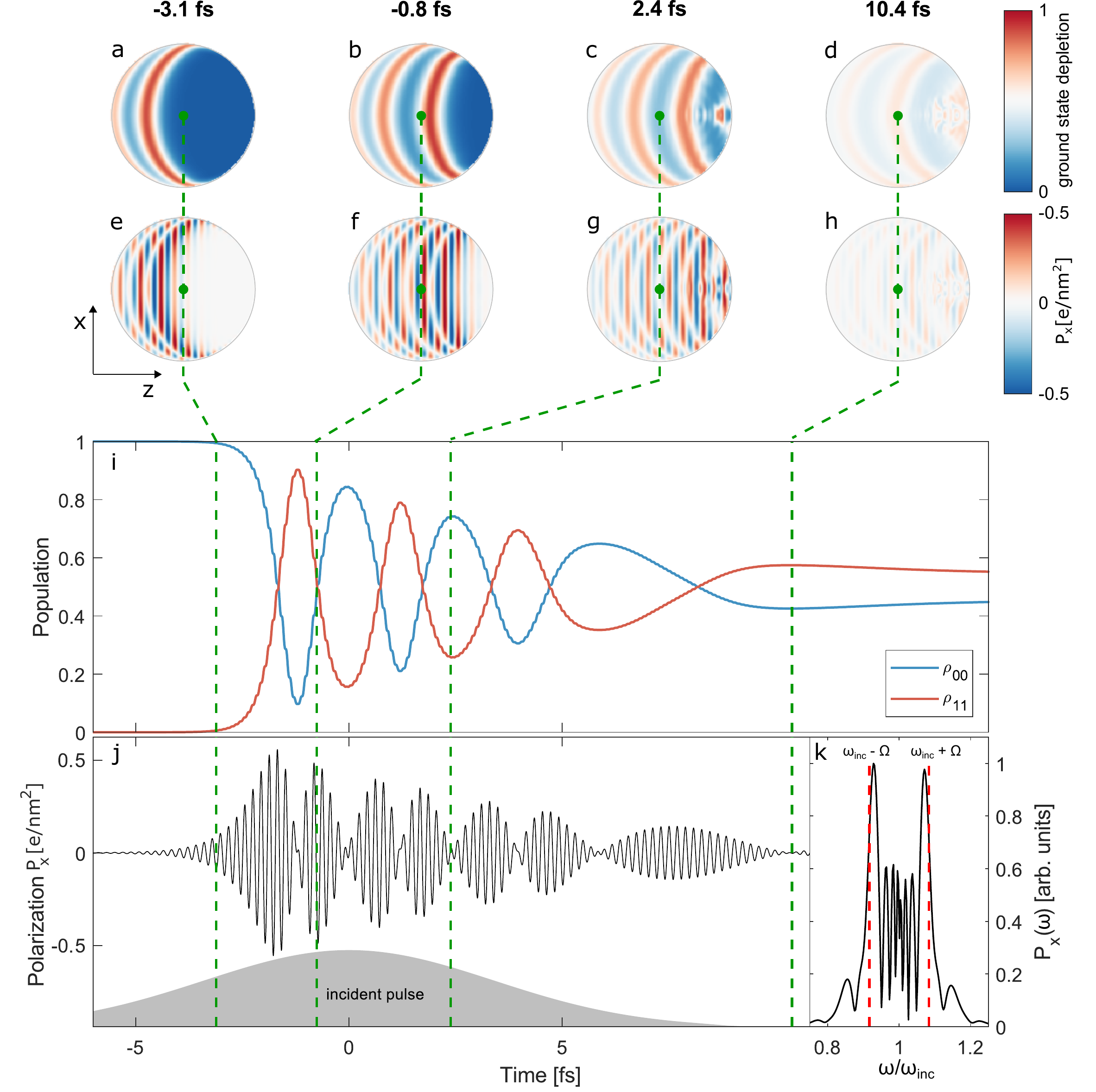}
	\end{center}
	\caption{Population and polarization dynamics in the He droplet for an excitation intensity of $5.0\times10^{14}\:\mathrm{W/cm^2}$. The upper two rows display snapshots of population (a-d) and  polarization (e-h) as cuts through the propagation-polarization plane. The detailed time-evolution of the population (i) and polarization (j) corresponds to the droplet center.	
{The gray shaded area indicates the incident free-space intensity envelope at the center plane of the droplet.} The spectrum of the polarization signal from panel (j) is shown in panel (k), documenting the spectral splitting and the formation of satellites that are up-/down shifted from the central frequency $\omega_{\mathrm{inc}}$ by roughly the Rabi frequency (as indicated). \label{fig:occ_dynamics}}
\end{figure}
The resulting snapshots of the ground state depletion in the propagation-polarization plane (Fig.~\ref{fig:occ_dynamics}a-d) display curved regions of similar depletion level that form excitation (or depletion) slabs whose structure approximately reflect the droplet curvature.  
The evolution of the depletion maps reveals traveling excitation waves (from left to right) with a focus-like convergence towards the back of the droplet. The corresponding snapshots of the local polarization (Fig.~\ref{fig:occ_dynamics}e-h) show that the phase fronts of the dipole excitation within the individual slabs are flat and do not reflect the curvature of the excitation slabs. However, the slab structure remains visible in the polarization signal in terms of the nodes of the polarization amplitude (white areas) that are associated with maxima and minima of the depletion (individual level population). 

The detailed evolution of the level population in the droplet center in Fig.~\ref{fig:occ_dynamics}(i) shows distinct signatures of Rabi-flopping between the ground and excited states $|0\rangle$  and $|1\rangle$ due to the coupling to the $x$-component of the electric field. Several Rabi-cycles are predicted before the coherent state dynamics ceases near the end of the pulse. Note that decoherence leads to a steady decrease of the amplitude of the oscillating population, underlining the necessity of short pulses for the realization of substantial Rabi cycling. The associated induced polarization in the droplet center displayed in Fig.~\ref{fig:occ_dynamics}(j) shows a clear fingerprint of the Rabi flopping and the associated dressed states, i.e. a signal modulation with the period of the Rabi cycle, with minimal amplitude for dominant population of only one state and maximal amplitude for equal population. The underlying field induced level splitting is expressed by the redistribution of the spectral polarization signal in two dominant satellites that are up-/downshifted from the central frequency of the incidence beam by the Rabi frequency $\Omega=\mu {\cal E}_0 $ where ${\cal E}_0$ is the laser electric field amplitude, see Fig.~\ref{fig:occ_dynamics}(k). The saturation effects, the decoherence, as well as the Rabi-cycling can be expected to cause substantial changes in the scattering images compared to the linear response result. In any case, at least for situations similar to the considered resonant scenario, the coherent state dynamics must obviously be taken into account already for intensities approaching $10^{14}\,{\rm W/cm^2}$, which are reachable in experiments already. 

\subsection{Nonlinear coherent diffractive imaging}
\begin{figure}[h]
\begin{center}
\includegraphics[width=1.0\textwidth]{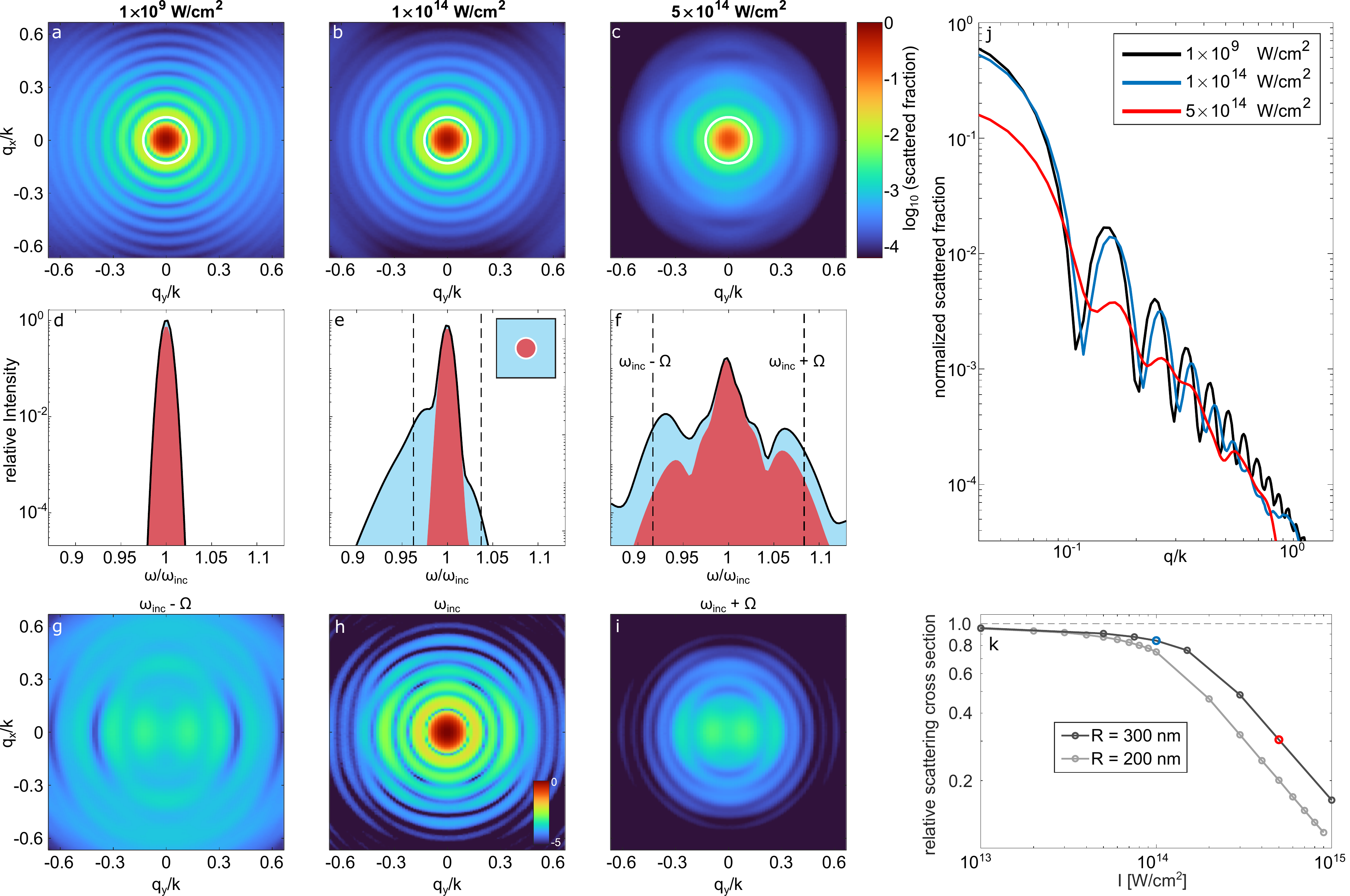}
\caption{Intensity dependence of the scattering patterns for spherical Helium droplets with radius $R=300$\,nm. Panels (a-c) show spectrally integrated scattering images (scattered fraction normalized to the center of the linear case) for three different excitation intensities (as indicated). The angular profiles in panel (j) show corresponding one-dimensional line-outs along the $q_x$ direction ($q_y=0$). Panels (d-f) show the associated spectra of the scattered fields (black solid lines) and their contributions for small and large scattering angles (red and blue shaded areas); the angle employed to discriminate small and large angle scattering is depicted as a white ring in panels (a-c). The bottom panels (g-i) show a spectrally selective analysis of the high intensity pattern in panel (c) representing the individual Fourier components at the central frequency $\omega_{\text{inc}}$ (h) and at the frequencies of the Rabi satellites $\omega_{\text{inc}}\mp \Omega$ (g, i). Panel (k) displays the intensity dependence of the scattering cross-section normalized to the linear response result for two droplet sizes (as indicated). \label{fig:nonlinear_cdi}}
\end{center}
\end{figure}

As the next step we investigate the structure and evolution of the scattering images in the regime of non-linear excitation. Figures~\ref{fig:nonlinear_cdi}(a-c) show three representative examples of spectrally integrated scattering patterns for different excitation intensities. 
Three main trends can be identified from the comparison. First, the total intensity of the scattering images shows a saturation behavior, as documented by the transition from a constant to a decreasing scattering cross section ($\propto1/I$) at high intensity, see Fig.~\ref{fig:nonlinear_cdi}(k). Note that the switching depends on the system size. Second, the spatial structure of the fringes becomes blurred with increasing intensity, exhibits a shift towards larger angles,  and shows a modified decay behavior, with reduced angular decay at high excitation intensity, see Fig.~\ref{fig:nonlinear_cdi}(j). Third, the spectra of the scattered fields widen substantially when compared with the linear-response results, see evolution of the full spectra (black line) in Figs.~\ref{fig:nonlinear_cdi}(d-f). The high intensity result shows clearly the emergence of an inelastically scattered radiation component in terms of satellites at  $\omega_{\text{inc}} \pm \Omega$ that is associated with the non-linear scattering process.

The saturation effect, as the most obvious feature, can be explained at least qualitatively from the fact that the polarization amplitude in the few-level system has an upper bound ($\propto \mu$). As a result, also the scattering from a pulse with finite duration must have an upper limit, impeding a linear scaling with the incidence pulse intensity at some point. The detailed modification of the fringe structure is less obvious. The dressed states created by the Rabi-cycling can be considered as an induced transparency, which will most strongly affect the droplet surface such that the effective sphere radius shrinks. This argument is  consistent with the slight increase of the fringe spacing in Fig.~\ref{fig:nonlinear_cdi}(j). For the blurring of the fringe features itself we consider two possible contributions. The first is the increased spectral width of the scattered fields. For linear scattering, spectral broadening reduces the fringe contrast in scattering patterns~\cite{SanderJPB2015} due to the mismatch in the fringe separations (Mie-scattering) associated with the different spectral contributions. However, another effect which we expect to be even more important for our case is connected with the structure of the excitation slabs. 

In order to motivate the slab structure effect we consider a slicing of the droplet into curved discs that resemble the typical thickness and curvature of the excitation slabs. The passage of the train of excitation waves through the spatially fixed slices leads to a the emission of a Rabi splitted spectrum. As individual slabs have different lateral sizes and thus create different fringe spacing in the diffraction pattern, the fringe structure in the full image is expected to blur if adjacent slices emit incoherently. This argument, however, would imply that the inelastically scattered radiation would show a weaker angular decay than the elastic (volume) contribution. The underlying logic is that a disc-like object shows a weaker angular decay in the scattering profile than a sphere. In the limit of the Born approximation the profiles of a disc and a sphere decay with~$q^{-3}$ and~$q^{-4}$, respectively. Though these values are modified in the presence of absorption, the trend of weaker angular decay for disc-like scatterers should remain. In fact, the frequency resolved, integrated scattering signal in Figs.~\ref{fig:nonlinear_cdi}(e,f) documents that the relative contribution of the inelastic signal is substantially higher for large scattering angles. Furthermore, the inelastic contribution shows a spectral asymmetry, which, however, also depends on scattering angle. In particular, in Fig.~\ref{fig:nonlinear_cdi}(f), the downshifted branch is dominant over the up-shifted feature for large scattering angles (blue shaded area), while the asymmetry is reversed for the small angle signal (red shaded area). A detailed analysis of this behavior, which also exhibits an intensity dependence,  is beyond the scope of the current work. 

Finally, if the inelastic signals of the scattered fields can be associated predominantly with the Rabi-cycling in the excitation-slabs, the spectrally selective analysis of the scattering images may provide rich information about the underlying spatio-temporal excitation dynamics. Figs.~\ref{fig:nonlinear_cdi}(g-i) display a corresponding analysis from the high intensity case in Figs.~\ref{fig:nonlinear_cdi}(c,f) and show the scattering images associated with only the scattered radiation at the central frequency and for frequencies around the two Rabi satellites, respectively. The deviation of the individual patterns, including the elastic one in Fig.~\ref{fig:nonlinear_cdi}(h), from the linear response result highlights the requirement to include the non-linear response in the analysis of scattering data.  Moreover, the fundamental structural differences between the individual patterns show that the non-linear response leaves a pronounced spectral fingerprint in the scattering patterns, opening a promising perspective for time-resolved characterization of the excitation dynamics.

\section{Conclusions}
\label{sec_Conclusions}
In summary, we developed a density-matrix-based description of coherent-diffractive imaging including quantum coherent bound-state dynamics and employed it to the scenario of resonance scattering from Helium nanodroplets. 
Although our model was used here for the describtion of a single resonance, it will be straightforward to describe more complex systems by adding additional active states. Our model is constructed such that it converges to the correct linear-response result for given optical parameters. For the investigated Helium droplets, our prediction at high intensity reveals a substantial impact of the non-linear quantum response. We find spatio-temporal excitation waves, Rabi-cycling, inelastic features in the scattered radiation, and pronounced deviations of the scattering patterns from the linear response case. Even though the presented application is based on strong approximations (single resonance, ionization neglected, correlated decay processes neglected), our findings show that quantum effects cannot be neglected for scenarios that are already experimentally in reach. The specific parameters of a corresponding experiment strongly depend on the particular objective. If coherent dynamics and Rabi satellites are of major interest, the laser pulse should be short enough to compete with the dephasing time but long enough to realize several Rabi cycles. If the goal is the mere observation of nonlinear features, the pulse duration is less critical and just the intensity has to be high enough to achieve substantial excited state population.
Finally, the rich features in the non-linear scattering reveal a promising potential for the characterization of complex quantum dynamics at the nanoscale via quantum coherent diffractive imaging and justify further in-depth analysis.

\ack
Financial support from the Deutsche Forschungsgemeinschaft via a Heisenberg-Grant (ID: 398382624) and via SPP1840 (ID: 281272685), from the Bundesministerium für Bildung und Forschung (BMBF, ID: 05K16HRB), and by the European Social Fund (ID: ESF/14-BM-A55-0007/19) and the Ministry of Education, Science and Culture of Mecklenburg-Vorpommern, Germany via project 'NEISS' is gratefully acknowledged. Computing time has been provided by the North German Supercomputing Alliance (HLRN, ID: mvp00013).

%\bibliography{Kruse_QCDI}

\providecommand{\newblock}{}

\end{document}